\documentclass[12pt]{article}
\usepackage{latexsym}
\usepackage{amsmath,amsfonts}
\usepackage{times}
\allowdisplaybreaks[4]

\catcode`\@=11

\newcount\hour
\newcount\minute
\newtoks\amorpm \hour=\time\divide\hour by 60\minute
=\time{\multiply\hour by 60 \global\advance\minute by-\hour}
\edef\standardtime{{\ifnum\hour<12 \global\amorpm={am}%
        \else\global\amorpm={pm}\advance\hour by-12 \fi
        \ifnum\hour=0 \hour=12 \fi
        \number\hour:\ifnum\minute<10
        0\fi\number\minute\the\amorpm}}
\edef\militarytime{\number\hour:\ifnum\minute<10 0\fi\number\minute}

\def\draftlabel#1{{\@bsphack\if@filesw {\let\thepage\relax
   \xdef\@gtempa{\write\@auxout{\string
      \newlabel{#1}{{\@currentlabel}{\thepage}}}}}\@gtempa
   \if@nobreak \ifvmode\nobreak\fi\fi\fi\@esphack}
        \gdef\@eqnlabel{#1}}
\def\@eqnlabel{}
\def\@vacuum{}
\def\marginnote#1{}
\def\draftmarginnote#1{\marginpar{\raggedright\scriptsize\tt#1}}
\def\draft{
        \pagestyle{plain}
        \overfullrule=2pt
        \oddsidemargin -.5truein
        \def\@oddhead{\sl \phantom{\today\quad\militarytime} \hfil
        \smash{\Large\sl DRAFT} \hfil \today\quad\militarytime}
        \let\@evenhead\@oddhead
        \let\label=\draftlabel
        \let\marginnote=\draftmarginnote
        \def\ps@empty{\let\@mkboth\@gobbletwo
        \def\@oddfoot{\hfil \smash{\Large\sl DRAFT} \hfil}
        \let\@evenfoot\@oddhead}
        \def\@eqnnum{(\theequation)\rlap{\kern\marginparsep\tt\@eqnlabel}%
        \global\let\@eqnlabel\@vacuum}  }

\newcommand{\rf}[1]{(\ref{#1})}
\renewcommand{\theequation}{\thesection.\arabic{equation}}
\renewcommand{\thefootnote}{\fnsymbol{footnote}}
\newcommand{\newsection}{   
\setcounter{equation}{0}\section}

\def\appendix#1{\addtocounter{section}{1}\setcounter{equation}{0}
\renewcommand{\thesection}{\Alph{section}}
\section*{Appendix \thesection\protect\indent \parbox[t]{11.15cm}{#1}}
\addcontentsline{toc}{section}{Appendix \thesection\ \ \ #1}}

\newcommand{\scsc}{\scriptscriptstyle}

\def\be{\begin{equation}}
\def\ee{\end{equation}}
\def\beq{\begin{eqnarray}}
\def\eeq{\end{eqnarray}}

\def\parline{\,\partial\kern -0.55em /\,\,}

\def\half{{\frac{1}{2}}}

\def\LL{{\cal L}}
\def\MM{{ M}}

\def\TT{{\cal T}}

\def\Nbf{{\bf N}}

\def\ibf{{\bf i}}
\def\iibf{{\bf ii}}

\def\alphab{{\bar{\alpha}}}

\def\rhob{{\bar{\rho}}}
\def\etab{{\bar{\eta}}}
\def\zetab{{\bar{\zeta}}}

\def\phik{|\phi\rangle}

\def\ck{|c\rangle}
\def\cbk{|\bar{c}\rangle}
\def\cbr{\langle c |}
\def\cbbr{\langle \bar{c}|}

\def\Phik{|\Phi\rangle}
\def\Phibr{\langle\Phi|}

\def\Xik{|\Xi\rangle}

\def\Ism{{\scriptscriptstyle I}}
\def\IIsm{{\scriptscriptstyle II}}
\def\FPsm{{\scriptscriptstyle FP}}

\def\Ab{\bar{A}}

\def\cb{\bar{c}}

\def\fb{\bar{f}}
\def\gb{\bar{g}}

\def\lb{\bar{l}}

\def\(A)dS{{\rm (A)dS}}

\def\I{{\rm I}}
\def\II{{\rm II}}

\def\ext{{\rm ext}}
\def\intrm{{\rm int}}
\def\gh{{\rm gh}}

\begin{document}


\begin{flushright}
FIAN-TD-2015-07 \qquad \ \ \ \ \ \ \  \\
arXiv: yymm.nnnn [hep-th] \\
\end{flushright}

\vspace{1cm}

\begin{center}

{\Large \bf BRST-BV approach to massless fields

\bigskip
adapted to AdS/CFT correspondence }

\vspace{2.5cm}

R.R. Metsaev%
\footnote{ E-mail: metsaev@lpi.ru
}

\vspace{1cm}

{\it Department of Theoretical Physics, P.N. Lebedev Physical
Institute, \\ Leninsky prospect 53,  Moscow 119991, Russia }

\vspace{3.5cm}

{\bf Abstract}

\end{center}

Using BRST-BV formulation of relativistic dynamics, arbitrary spin massless and massive field propagating in flat space and arbitrary spin massless fields propagating in AdS space are considered. For such fields, BRST-BV Lagrangians invariant under gauge transformations are obtained. The Lagrangians and gauge transformations are built in terms of traceless gauge fields and traceless gauge transformation parameters. The use of the Poincar\'e parametrization of AdS space allows us to get  simple BRST-BV Lagrangian for AdS fields. By imposing the Siegel gauge condition, we get gauge-fixed Lagrangian which leads to decoupled equations of motion for AdS fields. Such equations of motion considerably simplify the study of AdS/CFT correspondence. In the basis of conformal algebra, realization of relativistic symmetries on space of fields and antifields entering the BRST-BV formulation of AdS fields is obtained.

\newpage
\renewcommand{\thefootnote}{\arabic{footnote}}
\setcounter{footnote}{0}

\newsection{  Introduction }

BRST approach \cite{Becchi:1974xu} arisen as method that simplified analysis of the Slavnov-Taylor identities \cite{Slavnov:1972fg} in gauge field theories. At the present time, the BRST approach is the main method for studying ultraviolet divergencies and for building the relativistic  invariant and finite $S$-matrix in gauge field theories. Moreover, the various extended versions of the BRST approach, which involve the antifields, turn out to be not only powerful approaches for studying  quantum properties of quantized gauge field theories but also efficient approaches for  building classical gauge field theories \cite{Siegel:1984wx}. In this paper, it is the extended versions of the BRST approach that will be refereed to as BRST-BV formulation of relativistic dynamics.

On the one hand, theory of massless higher-spin fields in AdS space \cite{Vasiliev:1990en} is interesting candidate for the role of the theory of all fundamental interactions.
On the other hand, the duality conjecture by Maldacena triggered interest not only in studying the string/gauge theory dualities but also led to new interesting applications of the massless higher-spin field theory to the study of conformal field theory. Interrelation between field theories in AdS space and conformal field theories on the boundary of AdS space is governed by so called effective action. The effective action can be defined  through a continual integral over AdS fields with some special boundary conditions imposed on boundary values of AdS fields. The boundary values of AdS fields are identified with shadow fields, while the variational derivatives of the effective action with respect to the shadow fields are considered as correlation functions of the boundary conformal field theory, which, according to the duality conjecture, is dual to bulk theory of AdS fields. This is to say that
the problem of computation of the effective action is important for the study of AdS/CFT correspondence.

In view of importance for the study of AdS/CFT correspondence, the effective action has actively been investigated in the literature. In quadratic approximation, the tree-level effective action for massless spin-1, spin-2, and arbitrary spin massless AdS fields was computed in the respective Refs.\cite{Freedman:1998tz,Liu:1998bu,Metsaev:2009ym}, while the tree-level effective action for massive spin-1, spin-2 and arbitrary spin AdS fields was evaluated in the respective Refs.    \cite{Mueck:1998iz,Polishchuk:1999nh,Metsaev:2011uy} (see also Ref.\cite{Metsaev:2010zu}). Discussion of the group-theoretical issues related to the problem of computation of the effective action may be found in Refs.\cite{Dobrev:1998md}-\cite{Metsaev:2015rda}. Computation of $n$-point tree-level correlation functions for the massless higher-spin field theory was discussed in Refs.\cite{Didenko:2012tv} (see also interesting recent  Refs.\cite{Florakis:2014aaa}). Discussion of various aspects of one-loop quantum effective action may be found, e.g., in Refs.\cite{Tseytlin:2013jya}.

In Ref.\cite{Metsaev:2014vda}, we begun the study of the effective action by using the BRST approach. Namely, using the metric like formulation of arbitrary spin AdS fields and modified de Donder gauge condition found in Refs.\cite{Metsaev:2008ks,Metsaev:2009hp}, and applying the standard Faddeev-Popov procedure, we obtained simple action for free arbitrary spin AdS fields which is invariant under global BRST transformations and applied such action to the study of AdS/CFT correspondence. In the present paper, we obtain simple BRST-BV action for massless arbitrary spin AdS fields which is invariant under local gauge transformation and adapted to the study of AdS/CFT correspondence. Also we demonstrate that the action found in Ref.\cite{Metsaev:2014vda} can be obtained from the  BRST-BV action in this paper by using the Siegel gauge condition.

Our BRST-BV formulations of fields in $R^{d-1,1}$ and $AdS_{d+1}$ have many common features. Therefore, just to illustrate our approach, in Sec.\ref{sec-01}, we discuss BRST-BV formulation of totally symmetric arbitrary spin massless and massive fields propagating in flat space $R^{d-1,1}$. In Sec.3, using the Poincar\'e parametrization of AdS space, we obtain BRST-BV Lagrangian for totally symmetric arbitrary spin massless fields in $AdS_{d+1}$. Such Lagrangian, when using the Siegel gauge condition, leads to decoupled equations of motion which considerably simplify the study of AdS/CF correspondence. Also we discuss realization of relativistic symmetries of the $so(d,2)$ algebra on space of gauge fields and antifields entering BRST-BV approach.

\newsection{  Massless and massive fields in $R^{d-1,1}$  }\label{sec-01}

In this Section, just to illustrate our BRST-BV formulation, we consider totally symmetric arbitrary spin massless and massive fields propagating in flat space $R^{d-1,1}$.

\noindent {\bf Massless fields}. To describe field content entering the BRST-BV formulation we introduce Grassmann coordinate $\theta$, Grassmann even oscillators $\alpha^a$, and Grassmann odd oscillators $\eta$, $\rho$. The oscillators $\alpha^a$ transform as vector of the Lorentz algebra $so(d-1,1)$, while $\theta$ and the oscillators $\eta$, $\rho$ transform as scalars of the Lorentz algebra. Using $\theta$ and the oscillators, we introduce a ket-vector
\be \label{14082015-man-01}
\Phik = \Phi(x,\theta,\alpha,\eta,\rho)|0\rangle\,,
\ee
where $x$ stands for coordinates $x^a$ of space-time $R^{d-1,1}$. Note that, by definition, field $\Phi$ \rf{14082015-man-01} is Grassmann even. Usual fields depending on the space-time coordinates $x^a$ are obtained by expanding $\Phi$ \rf{14082015-man-01} into the Grassmann coordinate $\theta$ and the oscillators $\alpha^a$, $\eta$, $\rho$. For the description of massless spin-$s$ field, we impose the following algebraic constraints on ket-vector $\Phik$ \rf{14082015-man-01},
\beq
\label{14082015-man-02} && (N_\alpha + N_\eta + N_\rho - s)\Phik = 0 \,,
\\
\label{14082015-man-03} && \alphab^2\Phik=0\,.
\eeq
Definition of the operators $N_\alpha$, $N_\eta$, $N_\rho$, $\alphab^2$ may be found in Appendix (see relations \rf{22082015-man-01}, \rf{22082015-man-02}).
Constraint \rf{14082015-man-02} tells us that the ket-vector $\Phik$ is degree-$s$ homogeneous polynomial in the oscillators $\alpha^a$, $\eta$, $\rho$, while constraint \rf{14082015-man-03} implies that fields obtained by expanding the ket-vector $\Phi$ \rf{14082015-man-01} into the oscillators  $\alpha^a$ are traceless tensor fields of the Lorentz algebra $so(d-1,1)$.

To illustrate tensor fields entering ket-vector $\Phik$ \rf{14082015-man-01}, we note that the expansion of the ket-vector $\Phik$ into the Grassmann coordinate $\theta$ and the Grassmann odd oscillators  $\eta$, $\rho$ can be presented as
\beq
\label{15082015-man-01} \Phik & = & \phik + \theta |\phi_*\rangle\,,
\\
\label{15082015-man-02} && \phik = |\phi_\Ism\rangle + \rho |c\rangle + \eta |\cb\rangle + \rho \eta |\phi_\IIsm\rangle\,,
\\
\label{15082015-man-03} && |\phi_*\rangle = |\phi_{\Ism *}\rangle + \rho |\cb_*\rangle + \eta |c_*\rangle + \rho \eta |\phi_{\IIsm *}\rangle\,.
\eeq
Ket-vectors appearing on right hand side in \rf{15082015-man-02}, \rf{15082015-man-03} depend on the vector oscillators $\alpha^a$. Let us consider the ket-vectors appearing in \rf{15082015-man-02}. Taking into account \rf{14082015-man-02}, we find that ket-vectors \rf{15082015-man-02} depend on the oscillators $\alpha^a$ as
\beq
\label{15082015-man-04} && |\phi_\Ism\rangle = \frac{1}{s!}  \alpha^{a_1} \ldots \alpha^{a_s}  \phi_\Ism^{a_1\ldots a_s}(x)|0\rangle\,,
\\
\label{15082015-man-05} && \ck = \frac{1}{(s-1)!}  \alpha^{a_1} \ldots \alpha^{a_{s-1}}  c^{a_1\ldots a_{s-1}}(x)|0\rangle\,,
\\
\label{15082015-man-06} && \cbk = \frac{1}{(s-1)!}  \alpha^{a_1} \ldots \alpha^{a_{s-1}}  \cb^{a_1\ldots a_{s-1}}(x)|0\rangle\,,
\\
\label{15082015-man-07} && |\phi_\IIsm\rangle = \frac{1}{(s-2)!}  \alpha^{a_1} \ldots \alpha^{a_{s-2}}  \phi_\IIsm^{a_1\ldots a_{s-2}}(x)|0\rangle\,.
\eeq
From relations \rf{15082015-man-04}-\rf{15082015-man-07}, we find the field content which  enters the ket-vectors in \rf{15082015-man-02}. Namely, we see that there is one rank-$s$ tensor field denoted by $\phi_\Ism^{a_1\ldots a_s}$, two rank-$(s-1)$ tensor fields denoted by $c^{a_1\ldots a_{s-1}}$, $\cb^{a_1\ldots a_{s-1}}$, and one rank-$(s-2)$ tensor field denoted by  $\phi_\IIsm^{a_1\ldots a_{s-2}}$.
All just mentioned tensor fields are totally symmetric.
Constraint \rf{14082015-man-03} implies that all fields are the traceless tensors of the $so(d-1,1)$ algebra. Note also that representation of the ket-vectors of antifields \rf{15082015-man-03} in terms of the respective tensor antifields takes the same form as in   \rf{15082015-man-04}-\rf{15082015-man-07}. To summarize, our ket-vector $\Phik$ describes fields and antifields which are totally symmetric traceless tensors of the Lorentz algebra $so(d-1,1)$.

\noindent {\bf BRST-BV Lagrangian}. In the framework of BRST-BV approach, the general representation for gauge-invariant action of fields and antifields in $R^{d-1,1}$ takes the form \cite{Siegel:1984wx}
\be \label{15082015-man-09}
S = \int d^dx\, \LL\,, \hspace{1cm} \LL = \half \int d\theta \Phibr Q \Phik\,.
\ee
BRST operator $Q$ entering Lagrangian \rf{15082015-man-09} can be presented as
\be \label{15082015-man-10}
Q =  \theta (\Box - \MM^2) + M^{\eta a} \partial^a + M^\eta + M^{\eta\eta} \partial_\theta\,,
\ee
where $\Box=\partial^a\partial^a$ stands for the D'Alembert operator in $R^{d-1,1}$, while $\partial_\theta$ stands for left derivative of the Grassmann coordinate, $\partial_\theta = \partial/\partial\theta$. From   \rf{15082015-man-10}, we see that the BRST operator is entirely defined by the operators $\MM^2$, $M^{\eta a}$, $M^\eta$, $M^{\eta\eta}$. These operators depend on the oscillators and do not depend on the Grassmann coordinate $\theta$, the coordinates $x^a$ of space-time $R^{d-1,1}$, and the derivatives $\partial_\theta$, $\partial^a$.  The operator $M^2$ is square of the mass operator. The operators  $M^{\eta a}$, $M^\eta$, $M^{\eta\eta}$ will be referred to as spin operators in this paper.  The equation $Q^2=0$ leads to the following (anti)commutation relations for the spin operators and the operator $M^2$,
\beq
 \label{15082015-man-11} && \{ M^{\eta a},M^{\eta b}\} = - 2\eta^{ab} M^{\eta\eta}\,,
\\
\label{15082015-man-12} && \{ M^\eta ,M^\eta \} = 2 M^2 M^{\eta\eta}\,,
\\
\label{15082015-man-12-a1} && [M^2,M^{\eta a}]= 0\,, \hspace{1cm} [M^2,M^\eta]= 0\,, \hspace{1.2cm} [M^2,M^{\eta\eta}]= 0\,,
\\
\label{15082015-man-12-a2}
 && \{ M^{\eta a},M^\eta \}= 0\,, \qquad [M^{\eta a},M^{\eta\eta} ] = 0\,, \qquad [M^\eta,M^{\eta\eta}]= 0\,.
\eeq

Gauge symmetries of the action \rf{15082015-man-09} are given by
\be  \label{15082015-man-12-a3}
\delta \Phik = Q \Xik,
\ee
where the gauge transformation parameter ket-vector $\Xik$ depends on the Grassmann coordinate  $\theta$ and the oscillators $\alpha^a$, $\eta$, $\rho$. By definition, the ket-vector $\Xik$ satisfies the algebraic constraints,
\beq
\label{15082015-man-14} \Xik & = &  \Xi(x,\theta,\alpha,\eta,\rho)|0\rangle\,,
\\
\label{15082015-man-15} && (N_\alpha + N_\eta + N_\rho - s)\Xik = 0 \,,
\\
\label{15082015-man-16} && \alphab^2\Xik=0\,,
\eeq
where $\Xi$ \rf{15082015-man-14} is Grassmann odd. Comparing \rf{14082015-man-02}, \rf{14082015-man-03} and \rf{15082015-man-15}, \rf{15082015-man-16},  we see that the ket-vectors $\Xik$ and $\Phik$ satisfy the same algebraic constraints.  Therefore, taking into account our analysis of the algebraic constraints for the ket-vector $\Phik$, we conclude that the ket-vector $\Xik$ is built in terms of gauge transformation parameters which are totally symmetric traceless tensor fields of the Lorentz algebra $so(d-1,1)$. Also we note that the representation of the ket-vector $\Xik$ in terms of the tensor fields is obtained simply by replacing gauge fields in \rf{15082015-man-04}-\rf{15082015-man-07} by the  tensor fields of the gauge transformation parameters.

To summarize, the procedure for building the gauge-invariant action and the corresponding gauge transformations amounts to the problem for building the BRST operator. In turn, the procedure for building the BRST operator amounts to the problem for the building a realization for the spin operators which satisfy the (anti)commutation relations given in \rf{15082015-man-11}-\rf{15082015-man-12-a2}. In our approach, a massless field is described by ket-vector \rf{14082015-man-01} which satisfies the algebraic constraints
\rf{14082015-man-02}, \rf{14082015-man-03}. This implies that on space of ket-vector  \rf{14082015-man-01} we should realize the (anti)commutation relations given in \rf{15082015-man-11}-\rf{15082015-man-12-a2}. We find the following realization for the spin operators and the operator $M^2$
\beq
\label{15082015-man-17} &&  M^{\eta a} = \eta g_\rho\alphab^a + A^a \gb_\eta\etab\,,
\\
\label{15082015-man-18} && M^{\eta\eta} =\eta\etab\,,
\\
\label{15082015-man-19} && M^\eta = 0\,,
\\
\label{15082015-man-20} && M^2 =0 \,,
\eeq
where we use the notation
\beq
\label{15082015-man-21} && A^a \equiv \alpha^a - \alpha^2\frac{1}{2N_\alpha  + d} \alphab^a\,,
\\
\label{15082015-man-22} && g_\rho \equiv \Bigl(\frac{2s +d-4-2N_\rho}{2s +d-4}\Bigr)^{1/2} \,,
\qquad \gb_\eta \equiv - \Bigl(\frac{2s +d-4-2N_\eta}{2s +d-4}\Bigr)^{1/2}\,.
\eeq

Let us make the following comment. Constraint \rf{14082015-man-02} is well known in the literature. In the earlier literature, in place of constraint \rf{14082015-man-03}, the following constraint was used  (see, e.g.,Refs. \cite{Bengtsson:1987jt})
\be \label{14082015-man-03-a1}
(\alphab^2 + 2 \etab\rhob)\Phik=0\,.
\ee
When using \rf{14082015-man-03-a1}, relations \rf{15082015-man-18}-\rf{15082015-man-20} do not change, while the operator $M^{\eta a}$ takes the form $M^{\eta a} = \eta \alphab^a -\alpha^a \etab$. In this paper, we suggest to use the algebraic constraint \rf{14082015-man-03} because this constraint turns of to be convenient for developing the BRST-BV formulation of AdS field which is adapted to AdS/CFT correspondence. Study unconstrained BRST formulations may be found in Refs.\cite{Buchbinder:2001bs}.
Ignoring the traceless constraints for the ket-vector \rf{14082015-man-01} leads to BRST formulations for reducible higher-spin fields, which, in the literature, are sometimes referred to as triplets (see, e.g., Refs.\cite{Sagnotti:2003qa}).

\noindent {\bf Massive fields}. To develop the BRST-BV formulation of massive spin-$s$ field we introduce Grassmann coordinate $\theta$, Grassmann even oscillators $\alpha^a$, $\zeta$ and Grassmann odd oscillators $\eta$, $\rho$. The oscillators $\alpha^a$ transform as vector of the Lorentz algebra $so(d-1,1)$, while $\theta$ and the oscillators $\zeta$, $\eta$, $\rho$ transform as scalars of the Lorentz algebra. Using $\theta$ and the oscillators we introduce a ket-vector
\be \label{16082015-man-01}
\Phik = \Phi(x,\theta,\alpha,\zeta,\eta,\rho)|0\rangle\,,
\ee
which, by definition, satisfies the following algebraic constraints
\beq
\label{16082015-man-02} && (N_\alpha + N_\zeta + N_\eta + N_\rho - s)\Phik = 0 \,,
\\
\label{16082015-man-03} && \alphab^2\Phik=0\,.
\eeq
Also, by definition, the field $\Phi$ \rf{16082015-man-01} is Grassmann even.
Note that algebraic constraint \rf{16082015-man-03} takes the same form as the one for massless field in \rf{14082015-man-03}.

To illustrate tensor fields entering ket-vector $\Phik$ \rf{16082015-man-01}, we note that the expansion of the ket-vector $\Phik$ into the Grassmann coordinate $\theta$ and the Grassmann odd oscillators  $\eta$, $\rho$ takes the same form as in \rf{15082015-man-01}-\rf{15082015-man-03}. Note however that, for massive fields, the ket-vectors appearing on the right hand side in \rf{15082015-man-02}, \rf{15082015-man-03} depend not only on the vector oscillators $\alpha^a$ but also on the scalar oscillator $\zeta$. For example, consider  the ket-vectors appearing in \rf{15082015-man-02}. Taking into account  \rf{16082015-man-02}, we find that ket-vectors \rf{15082015-man-02} depend on the oscillators $\alpha^a$, $\zeta$ as
\beq
\label{16082015-man-04} && |\phi_\Ism\rangle = \sum_{s'=0}^s \frac{\zeta^{s-s'}}{\sqrt{(s-s')!}} |\phi_\Ism^{s'}\rangle \,, \hspace{2.3cm} |\phi_\Ism^{s'}\rangle = \frac{1}{s'!}  \alpha^{a_1} \ldots \alpha^{a_{s'}}  \phi_\Ism^{a_1\ldots a_{s'}}(x)|0\rangle\,,
\\
\label{16082015-man-05} && \ck = \sum_{s'=0}^{s-1} \frac{\zeta^{s-1-s'}}{\sqrt{(s-1-s')!}} |c^{s'}\rangle \,, \hspace{1.8cm} |c^{s'}\rangle = \frac{1}{s'!}  \alpha^{a_1} \ldots \alpha^{a_{s'}}  c^{a_1\ldots a_{s'}}(x)|0\rangle\,,
\\
\label{16082015-man-06} && \cbk = \sum_{s'=0}^{s-1} \frac{\zeta^{s-1-s'}}{\sqrt{(s-1-s')!}} |\cb^{s'}\rangle \,, \hspace{1.9cm}  |\cb^{s'}\rangle = \frac{1}{s'!}  \alpha^{a_1} \ldots \alpha^{a_{s'}}  \cb^{a_1\ldots a_{s'}}(x)|0\rangle\,,
\\
\label{16082015-man-07} && |\phi_\IIsm\rangle = \sum_{s'=0}^{s-2} \frac{\zeta^{s-2-s'}}{\sqrt{(s-2-s')!}} |\phi_\IIsm^{s'}\rangle \,, \hspace{1.4cm} |\phi_\IIsm^{s'}\rangle = \frac{1}{s'!}  \alpha^{a_1} \ldots \alpha^{a_{s'}}  \phi_\IIsm^{a_1\ldots a_{s'}}(x)|0\rangle\,.\qquad
\eeq
For massive field, ket-vectors appearing on the right hand side in \rf{15082015-man-03} can be represented in terms of tensor antifields in the same way as in \rf{16082015-man-04}-\rf{16082015-man-07}.

To build BRST operator \rf{15082015-man-10} we should find the spin operators which satisfy the (anti)com\-mutation relations \rf{15082015-man-11}-\rf{15082015-man-12-a2}. In our approach, the massive field is described by  ket-vector \rf{16082015-man-01} which satisfies the algebraic constraints
\rf{16082015-man-02}, \rf{16082015-man-03}. Thus, we should find a realization of the spin operators on space of ket-vector \rf{16082015-man-01}. We find the following realization for the spin operators and the operator $M^2$
\beq
\label{16082015-man-08} &&  M^{\eta a} = \eta g_{\rho \zeta}\alphab^a + A^a \gb_{\eta \zeta}\etab\,,
\\
\label{16082015-man-09} && M^{\eta\eta} =\eta\etab\,,
\\
\label{16082015-man-10} && M^\eta = \eta l_{\rho \zeta}\zetab + \zeta \lb_{\eta \zeta}\etab \,,
\\
\label{16082015-man-11} && M^2 = m^2 \,,
\eeq
where $m$ in \rf{16082015-man-11} stands for mass parameter and we use the following notation
\beq
\label{16082015-man-14} && A^a \equiv \alpha^a - \alpha^2\frac{1}{2N_\alpha  + d} \alphab^a\,,
\\
\label{16082015-man-15} && g_{\rho \zeta} \equiv \Bigl(\frac{2s +d-4-2N_\zeta-2N_\rho}{2s +d-4-2N_\zeta}\Bigr)^{1/2} \,,
\\
\label{16082015-man-16} && \gb_{\eta \zeta} \equiv - \Bigl(\frac{2s +d-4-2N_\zeta-2N_\eta}{2s +d-4-2N_\zeta}\Bigr)^{1/2}\,,
\\
\label{16082015-man-17} && l_{\rho \zeta} \equiv m e_\zeta\Bigl(\frac{2s +d-4-2N_\zeta-2N_\rho}{2s +d-4-2N_\zeta}\Bigr)^{-1/2} \,,
\\
\label{16082015-man-18} && \lb_{\eta \zeta} \equiv m e_\zeta\Bigl(\frac{2s +d-4-2N_\zeta-2N_\eta}{2s +d-4-2N_\zeta}\Bigr)^{-1/2}\,,
\\
\label{16082015-man-19} && \hspace{1cm} e_\zeta \equiv \Bigl(\frac{2s +d-4-N_\zeta}{2s +d-4-2N_\zeta}\Bigr)^{1/2}\,.
\eeq
Gauge symmetries of action \rf{15082015-man-09} take the same form as in \rf{15082015-man-12-a3},
where, in place of ket-vector $\Phik$, we should use the ket-vector $\Phik$ given in \rf{16082015-man-01}, while, in place of gauge transformation parameter ket-vector $\Xik$, we should use a ket-vector $\Xik$ which is  appropriate to massive field. Namely, in the case under consideration, the ket-vector $\Xik$ depends on the Grassmann coordinate $\theta$ and the oscillators $\alpha^a$, $\zeta$, $\eta$, $\rho$. Also, the ket-vector $\Xik$ should satisfy the constraints which take the same form as the ones for the ket-vector $\Phik$ in \rf{16082015-man-02}, \rf{16082015-man-03},
\beq
\label{16082015-man-19-a1} \Xik & = &  \Xi(x,\theta,\alpha,\zeta,\eta,\rho)|0\rangle\,,
\\
\label{16082015-man-19-a2} && (N_\alpha + N_\zeta + N_\eta + N_\rho - s)\Xik = 0 \,, \qquad \alphab^2\Xik=0\,.\qquad
\eeq

\newsection{  Massless fields in $AdS_{d+1}$ }

In this Section, we develop BRST-BV formulation of totally symmetric spin-$s$ massless field in $AdS_{d+1}$ adapted to AdS/CFT correspondence. To this end we use the Poincar\'e parametrization of $AdS_{d+1}$,
\be \label{16082015-man-20}
ds^2 = \frac{1}{z^2}(dx^a dx^a + dz dz)\,, \qquad a=0,1,\ldots,d-1\,.
\ee
From \rf{16082015-man-20}, we see that, when using the Poincar\'e parametrization of $AdS_{d+1}$, only symmetries of the $so(d-1,1)$ algebra are manifest.

\noindent {\bf Field content}. To describe field content entering the BRST-BV formulation we introduce Grassmann coordinate $\theta$, Grassmann even oscillators $\alpha^a$, $\alpha^z$, and Grassmann odd oscillators $\eta$, $\rho$. The oscillators $\alpha^a$ transform as vector of the $so(d-1,1)$ algebra, while $\theta$ and the oscillators $\alpha^z$, $\eta$, $\rho$ transform as scalars of the $so(d-1,1)$ algebra. Using $\theta$ and the oscillators, we introduce a ket-vector
\be \label{16082015-man-21}
\Phik = \Phi(x,z,\theta,\alpha,\alpha^z,\eta,\rho)|0\rangle\,,
\ee
where $x$, $z$ stand for the Poincar\'e coordinates $x^a$, $z$ of $AdS_{d+1}$ space \rf{16082015-man-20}. By definition, field $\Phi$ \rf{16082015-man-21} is Grassmann even. Usual tensor fields depending on the space-time coordinates $x^a$, $z$ appear by expanding ket-vector $\Phik$ \rf{16082015-man-21} into the Grassmann coordinate $\theta$ and the oscillators $\alpha^a$, $\alpha^z$, $\eta$, $\rho$. To discuss the spin-$s$ massless field we impose the following algebraic constraints on ket-vector $\Phik$ \rf{16082015-man-21},
\beq
 \label{16082015-man-22} && (N_\alpha + N_{\alpha^z} +  N_\eta + N_\rho - s)\Phik = 0 \,,
\\
 \label{16082015-man-23} && \alphab^2\Phik=0\,.
\eeq
Definition of the operators $N_\alpha$, $N_z$, $N_\eta$, $N_\rho$, $\alphab^2$ may be found in Appendix (see \rf{22082015-man-01}, \rf{22082015-man-02}). Constraint \rf{16082015-man-22} implies  that ket-vector $\Phik$ \rf{16082015-man-21} is degree-$s$ homogeneous polynomial in the oscillators $\alpha^a$, $\alpha^z$, $\eta$, $\rho$, while constraint \rf{16082015-man-23} tells us that fields which are obtained by expanding the ket-vector  $\Phik$ \rf{16082015-man-21} into the oscillators $\alpha^a$, are traceless tensor fields of the $so(d-1,1)$ algebra.

We now illustrate tensor fields entering ket-vector $\Phik$ \rf{16082015-man-21}. First, we note that the expansion of the ket-vector $\Phik$ into the Grassmann coordinate $\theta$ and the Grassmann odd oscillators  $\eta$, $\rho$ takes the same form as in \rf{15082015-man-01}-\rf{15082015-man-03}.
Second, the algebraic constraint \rf{16082015-man-22} for massless field in $AdS_{d+1}$ is obtained from the one for massive field $R^{d-1,1}$, \rf{16082015-man-02}, by  making the substitution of the scalar oscillators $\zeta \rightarrow \alpha^z$. This implies that, for massless AdS field, ket-vectors  \rf{15082015-man-02} are obtained by making the substitution $\zeta \rightarrow \alpha^z$ in relations \rf{16082015-man-04}-\rf{16082015-man-07},

\beq
\label{16082015-man-24} && |\phi_\Ism\rangle = \sum_{s'=0}^s \frac{\alpha_z^{s-s'}}{\sqrt{(s-s')!}} |\phi_\Ism^{s'}\rangle \,, \hspace{2.3cm} |\phi_\Ism^{s'}\rangle = \frac{1}{s'!}  \alpha^{a_1} \ldots \alpha^{a_{s'}}  \phi_\Ism^{a_1\ldots a_{s'}}(x,z)|0\rangle\,,
\\
\label{16082015-man-25} && \ck = \sum_{s'=0}^{s-1} \frac{\alpha_z^{s-1-s'}}{\sqrt{(s-1-s')!}} |c^{s'}\rangle \,, \hspace{1.8cm} |c^{s'}\rangle = \frac{1}{s'!}  \alpha^{a_1} \ldots \alpha^{a_{s'}}  c^{a_1\ldots a_{s'}}(x,z)|0\rangle\,,
\\
\label{16082015-man-26} && \cbk = \sum_{s'=0}^{s-1} \frac{\alpha_z^{s-1-s'}}{\sqrt{(s-1-s')!}} |\cb^{s'}\rangle \,, \hspace{1.9cm}  |\cb^{s'}\rangle = \frac{1}{s'!}  \alpha^{a_1} \ldots \alpha^{a_{s'}}  \cb^{a_1\ldots a_{s'}}(x,z)|0\rangle\,,
\\
\label{16082015-man-27} && |\phi_\IIsm\rangle = \sum_{s'=0}^{s-2} \frac{\alpha_z^{s-2-s'}}{\sqrt{(s-2-s')!}} |\phi_\IIsm^{s'}\rangle \,, \hspace{1.4cm} |\phi_\IIsm^{s'}\rangle = \frac{1}{s'!}  \alpha^{a_1} \ldots \alpha^{a_{s'}}  \phi_\IIsm^{a_1\ldots a_{s'}}(x,z)|0\rangle\,.\qquad
\eeq

For massless field in $AdS_{d+1}$, ket-vectors of antifields \rf{15082015-man-03} involved in the BRST-BV formulation can be represented in terms of tensor antifields in the same way as in \rf{16082015-man-24}-\rf{16082015-man-27}.

\noindent {\bf BRST-BV Lagrangian}. In the framework of BRST-BV approach, the general representation for the gauge invariant action we find takes the form

\be \label{16082015-man-29}
S = \int d^dx dz\, \LL\,, \hspace{2cm} \LL = \half \int d\theta \Phibr Q \Phik\,.
\ee
BRST operator $Q$ in \rf{16082015-man-29} admits the following representation

\be \label{16082015-man-30}
Q =  \theta (\Box - \MM^2) + M^{\eta a} \partial^a + M^\eta + M^{\eta\eta} \partial_\theta\,,
\ee
where $\Box=\partial^a\partial^a$ stands for the D'Alembert operator in $R^{d-1,1}$, while  $\partial_\theta$ stands for left derivative of the Grassmann coordinate,  $\partial_\theta = \partial/\partial\theta$. From \rf{16082015-man-30}, we see that the BRST operator is defined entirely in terms of the operators $\MM^2$, $M^{\eta a}$, $M^\eta$, $M^{\eta\eta}$. All these operators depend on the oscillators. The operators $M^{\eta a}$, $M^{\eta\eta}$ do not depend on the Grassmann coordinate   $\theta$, the coordinate of AdS space, $x^a$, $z$, and the derivatives $\partial_\theta$, $\partial^a$, $\partial_z$. The operators $M^2$, $M^\eta$ also do not depend on the Grassmann coordinate $\theta$, the boundary coordinates $x^a$, and the derivatives $\partial_\theta$, $\partial^a$. However the operators $M^2$, $M^\eta$  depend on the radial coordinate $z$ and the radial derivative $\partial_z$. Equation $Q^2=0$ leads to (anti)commutation relations given in \rf{15082015-man-11}-\rf{15082015-man-12-a2}.

Comparing \rf{15082015-man-09}, \rf{15082015-man-10} and \rf{16082015-man-29}, \rf{16082015-man-30}, we note that the general structure of the BRST-BV action we suggest in this paper for the description of fields in $AdS_{d+1}$ is similar to the one for fields in flat space $R^{d-1,1}$ suggested in Refs.\cite{Siegel:1984wx}. We note that, it is the use of the Poincar\'e parametrization of AdS space \rf{16082015-man-20} that allows us to cast the BRST-BV action into the form presented in  \rf{16082015-man-29}, \rf{16082015-man-30}.

To build BRST operator $Q$ \rf{16082015-man-30} we should find the spin operators which satisfy the (anti)com\-mutation relations in \rf{15082015-man-11}-\rf{15082015-man-12-a2}. In our approach, the massless field is described by ket-vector \rf{16082015-man-21} which satisfies the algebraic constraints \rf{16082015-man-22}, \rf{16082015-man-23}. This is to say that we should build a realization of the spin operators on space of ket-vector \rf{16082015-man-21}. We find the following realization for the spin operators and the operator $M^2$

\beq
\label{16082015-man-31} &&  M^{\eta a} = \eta g_{\rho z}\alphab^a + A^a \gb_{\eta z}\etab\,,
\\
\label{16082015-man-32} && M^{\eta\eta} =\eta\etab\,,
\\
\label{16082015-man-33} && M^\eta = \eta l_{\rho z}\alphab^z + \alpha^z \lb_{\eta z}\etab \,,
\\
\label{16082015-man-34} && M^2 = - \partial_z^2 + \frac{1}{z^2}(\nu^2 - \frac{1}{4}) \,,
\eeq
where the operators $A^a$, $g_{\rho z}$, $\gb_{\eta z}$, $l_{\rho z}$, $\lb_{\eta z}$, $\nu$ are defined by the relations
\beq
\label{16082015-man-35} && A^a \equiv \alpha^a - \alpha^2\frac{1}{2N_\alpha  + d} \alphab^a\,,
\\
\label{16082015-man-36} && g_{\rho z} \equiv \Bigl(\frac{2s +d-4-2N_z-2N_\rho}{2s +d-4-2N_z}\Bigr)^{1/2} \,,
\\
\label{16082015-man-37} && \gb_{\eta z} \equiv - \Bigl(\frac{2s +d-4-2N_z-2N_\eta}{2s +d-4-2N_z}\Bigr)^{1/2}\,,
\\
\label{16082015-man-38} && l_{\rho z} \equiv \TT_{-\nu+\half} e_z\Bigl(\frac{2s +d-4-2N_z-2N_\rho}{2s +d-4-2N_z}\Bigr)^{-1/2} \,,
\\
\label{16082015-man-39} && \lb_{\eta z} \equiv -  \TT_{\nu-\half} e_z\Bigl(\frac{2s +d-4-2N_z-2N_\eta}{2s +d-4-2N_z}\Bigr)^{-1/2}\,,
\\
\label{16082015-man-40} && \hspace{1cm} \nu \equiv s + \frac{d-4}{2} - N_z\,,
\\
\label{16082015-man-41} && \hspace{1cm} e_z \equiv \Bigl(\frac{2s +d-4-N_z}{2s +d-4-2N_z}\Bigr)^{1/2}\,,
\\
\label{16082015-man-42} &&  \hspace{1cm} \TT_\nu \equiv \partial_z + \frac{\nu}{z}\,.
\eeq

For massless AdS field, gauge symmetries of the action \rf{16082015-man-29}  take the same form as in \rf{15082015-man-12-a3} where, in place of ket-vector $\Phik$, we should use the ket-vector $\Phik$ given in \rf{16082015-man-21}, while, in place of gauge transformation parameter ket-vector $\Xik$, we should use a ket-vector $\Xik$ which is appropriate to massless AdS field. Namely, in the case under consideration, the ket-vector $\Xik$ depends on the Grassmann coordinate $\theta$ and the oscillators $\alpha^a$, $\alpha^z$, $\eta$, $\rho$. Also, the ket-vector $\Xik$ should satisfy the constraints which take the same form as ones for the ket-vector $\Phik$ in \rf{16082015-man-22}, \rf{16082015-man-23},
\beq
\label{16082015-man-19-b1} \Xik & = &  \Xi(x,z,\theta,\alpha,\alpha^z,\eta,\rho)|0\rangle\,,
\\
\label{16082015-man-19-b2} && (N_\alpha + N_z + N_\eta + N_\rho - s)\Xik = 0 \, \qquad \alphab^2\Xik=0\,. \qquad
\eeq

We make the following two comments.

\noindent \ibf) From relations \rf{16082015-man-34}, \rf{16082015-man-40}, we see that the operator $M^2$ is diagonal on space of the ket-vector $\Phik$. It is this property of the operator $M^2$ that seems to be most attractive feature of our approach. This is to say that it is the diagonal representation for the operator $M^2$ that allows us in an easy way to solve the equations of motion for AdS field, compute the effective action and hence to study AdS/CFT correspondence. For example, using the Siegel gauge $|\phi_*\rangle = 0$ and integrating over the Grassmann coordinate $\theta$, we verify that Lagrangian \rf{16082015-man-29} takes the form
\be \label{18082015-man-02}
\LL =  \half \langle \phi_{_\I} | (\Box - M^2)  |\phi_{_\I} \rangle - \half \langle \phi_{_\II} | (\Box - M^2) |\phi_{_\II} \rangle
+  \langle \cb |(\Box - M^2)  |c\rangle\,.
\ee
For the first time, Lagrangian \rf{18082015-man-02} was obtained in Ref.\cite{Metsaev:2014vda} by using the standard Faddeev-Popov procedure. Obviously, Lagrangian \rf{18082015-man-02} leads to decoupled equations of motion and this considerably simplifies the study of AdS/CFT correspondence (see Ref.\cite{Metsaev:2014vda}). Note that, when passing from $\LL$  \rf{16082015-man-29} to $\LL$ \rf{18082015-man-02}, we made the substitution $\cbk\rightarrow -\cbk$.

\noindent \iibf) In Ref.\cite{Metsaev:1999ui}, using light-cone gauge, we demonstrated that arbitrary spin massless field in $AdS_{d+1}$ can be presented as direct sum of massive fields $R^{d-1,1}$ with continuous mass spectrum. The BRST-BV formulation we developed in this paper allows us to demonstrate this fact for gauge massless field in $AdS_{d+1}$ and massive fields in $R^{d-1,1}$ in a straightforward way. To this end we use the new notation $|\Phi_{\scsc AdS_{d+1}}(x,z)\rangle$ for the ket-vector of massless AdS field $\Phik$ \rf{16082015-man-21}.
The ket-vector of massive field $\Phik$ \rf{16082015-man-01}, after the substitution $\zeta\rightarrow \alpha^z$, will be denoted as $|\Phi_{\scsc R^{d-1,1} }(x,m)\rangle$. We now suggest the following interrelation for the ket-vectors $|\Phi_{\scsc AdS_{d+1}}(x,z)\rangle$ and $|\Phi_{\scsc R^{d-1,1} }(x,m)\rangle$ by using the Bessel transformation with respect to the radial coordinate $z$,
\be \label{18082015-man-03}
|\Phi_{AdS_{d+1}}(x,z)\rangle = \int\limits_0^\infty dm\, Z_\nu(m z) (-)^{N_z}|\Phi_{R^{d-1,1}}(x,m)\rangle\,, \hspace{1cm} Z_\nu(z) \equiv \sqrt{z}J_\nu(z)\,.
\ee
We prove relation \rf{18082015-man-03} in the following way. We use the notation $G_{\scsc AdS_{d+1} }$ for the operators given in \rf{16082015-man-31}-\rf{16082015-man-34}. After the substitution $\zeta\rightarrow \alpha^z$ in \rf{16082015-man-08}-\rf{16082015-man-11}, the operators given  \rf{16082015-man-08}-\rf{16082015-man-11} will be denoted by $G_{\scsc R^{d-1,1} }$. Now, in order to prove interrelation \rf{18082015-man-03} we should verify that the following relation

\be \label{18082015-man-04}
G_{\scsc AdS_{d+1} } |\Phi_{\scsc AdS_{d+1}}(x,z)\rangle = \int  dm\, Z_\nu(m z) (-)^{N_z} G_{\scsc R^{d-1,1} } |\Phi_{\scsc R^{d-1,1}}(x,m)\rangle
\ee
holds true.
Operators $M^{\eta a}$, $M^{\eta\eta}$ given in \rf{16082015-man-08}, \rf{16082015-man-09} and \rf{16082015-man-31},\rf{16082015-man-32}
obviously satisfy the relation \rf{18082015-man-04}.  For operators $M^\eta$, $M^2$ given in  \rf{16082015-man-10}, \rf{16082015-man-11} and \rf{16082015-man-33},\rf{16082015-man-34}, the relations \rf{18082015-man-04} can easily be proved by using the following identities for the function $Z_\nu$ defined in \rf{18082015-man-03}

\be
\alpha^z \TT_{\nu-\half} Z_\nu  = Z_\nu \alpha^z \,, \quad  \TT_{-\nu+\half} \alphab^z Z_\nu = - Z_\nu \alphab^z\,, \quad \TT_{\nu-\half} Z_\nu  = Z_{\nu-1} \,, \quad  \TT_{-\nu-\half} Z_\nu = - Z_{\nu+1}\,,
\ee
where the operators $\nu$, $\TT_\nu$ are defined in \rf{16082015-man-40}, \rf{16082015-man-42}. Note that operator $M^2$ \rf{16082015-man-34} can be represented as $M^2 = - \TT_{\scsc -\nu + \half} \TT_{\scsc \nu-\half}$.

\medskip
\noindent {\bf Realization of relativistic symmetries}. Relativistic symmetries of fields in  $AdS_{d+1}$ are described by the $so(d,2)$ algebra. In our approach, only symmetries of the $so(d-1,1)$ algebra are manifest. Therefore we should build the realizations of symmetries of the   $so(d,2)$ algebra on space of gauge fields and antifields entering the ket-vector $\Phik$ \rf{16082015-man-21}. As the symmetries of the $so(d-1,1)$ algebra are manifest in our approach, it is reasonable to represent the generators of the $so(d,2)$ algebra in the basis of the $so(d-1,1)$ algebra. In such basic, sometimes to be referred to as conformal algebra basis, the $so(d,2)$ algebra is given by the translations $P^a$, dilatation $D$, conformal boosts $K^a$, and generators of the  $so(d-1,1)$ algebra, $J^{ab}$. We use the following commutation relations for the generators

\beq
&& {}[D,P^a]=-P^a\,,
\hspace{2.5cm}
[P^a,J^{bc}]
=\eta^{ab}P^c
-\eta^{ac}P^b
\,,
\nonumber\\
&& [D,K^a]=K^a\,,
\hspace{2.7cm}
[K^a,J^{bc}]
=\eta^{ab}K^c
- \eta^{ac}K^b\,,
\\
&& [P^a,K^b]
= \eta^{ab}D
- J^{ab}\,,
\hspace{1.2cm}
[J^{ab},J^{ce}]
= \eta^{bc} J^{ae}
+ 3\hbox{ terms}.
\nonumber
\eeq
On space of fields and antifields entering  the BRST-BV approach,
we find the following general realization for generators of the $so(d,2)$ algebra
\beq
\label{16082015-man-46} && P^a = \partial^a\,, \hspace{1cm} J^{ab} = x^a\partial^b - x^b \partial^a + M^{ab}\,,
\hspace{1cm} D = x^a\partial^a + \Delta\,,
\nonumber\\
\label{16082015-man-47} && K^a =  - \half x^2 \partial^a + x^a D + M^{ab} x^b + R^a\,, \qquad
\\
\label{16082015-man-48} && \hspace{1cm} \Delta \equiv z\partial_z + 2\theta\partial_\theta + M^{\eta\rho} + \frac{d-1}{2}\,,
\\
\label{16082015-man-49} && \hspace{1cm} R^a = - \half z^2 \partial^a - M^{za} z + 2 \theta M^{\rho a}\,.
\eeq
Operators $M^{ab}$, $M^{za}$, $M^{\eta\rho}$, $M^{\rho a}$ \rf{16082015-man-46}-\rf{16082015-man-49} do not depend on the Grassmann coordinate $\theta$, the coordinates of AdS space $x^a$, $z$, and the derivatives $\partial_\theta$, $\partial^a$, $\partial_z$. These operators depend only on the oscillators. On space of ket-vector $\Phik$ \rf{16082015-man-21}, we find the following realization for the operators $M^{ab}$, $M^{za}$, $M^{\eta\rho}$, $M^{\rho a}$,
\beq
&& M^{ab} = \alpha^a\alphab^b - \alpha^b \alphab^a\,,
\\
\label{16082015-man-50} && M^{za} = A^z \alphab^a + A^a \Ab^z\,,
\\
\label{16082015-man-51} && M^{\eta\rho} = N_\eta - N_\rho\,,
\\
\label{16082015-man-52} && M^{\rho a} = \rho f_{\eta z}\alphab^a + A^a \fb_{\rho z}\rhob\,,
\eeq
where the operator $A^a$ is defined by relation \rf{16082015-man-35}. In \rf{16082015-man-50}-\rf{16082015-man-52}, we use the following notation
\beq
\label{16082015-man-53} && A^z \equiv \alpha^z e_{\eta\rho z} + h_z \rho \eta \alphab^z\,,
\nonumber\\
\label{16082015-man-54} &&  \Ab^z \equiv - e_{\eta\rho z}\alphab^z - \alpha^z\etab\rhob h_z\,,
\nonumber\\
\label{16082015-man-55} && f_{\rho z} \equiv \Bigl(\frac{2s +d-4-2N_z-2N_\rho}{2s +d-4-2N_z}\Bigr)^{1/2} \,,
\nonumber\\
\label{16082015-man-56} && \fb_{\eta z} \equiv  \Bigl(\frac{2s +d-4-2N_z-2N_\eta}{2s +d-4-2N_z}\Bigr)^{1/2}\,,
\nonumber\\
\label{16082015-man-57} && e_{\eta\rho z} \equiv e_z\Bigl(\frac{(2s +d-4-2N_z)(2s +d-4-2N_z-2N_\eta - 2N_\rho)}{(2s + d-4-2N_z-2N_\eta)(2s +d-4-2N_z-2N_\rho)}\Bigr)^{1/2} \,,
\nonumber\\
\label{16082015-man-59} && h_z \equiv 2e_z \bigl((2s +d-4-2N_z)(2s +d-6-2N_z)\bigr)^{-1/2}\,,
\eeq
where the operator $e_z$ is defined in \rf{16082015-man-41}.

\bigskip
\noindent {\bf Symmetries of the $osp(d-1,1|2)$ superalgebra}. From relations \rf{16082015-man-33}, \rf{16082015-man-34}, we see that the operators $M^2$, $M^\eta$ depend not only on the oscillators but also on the radial coordinate $z$ and the radial derivative $\partial_z$. Therefore, on the one hand, consider the spin operators $M^{\eta a}$, $M^{\eta \eta}$, \rf{16082015-man-31}, \rf{16082015-man-32}, which depend only on the oscillators. On the other hand, from the expressions for the generators of the relativistic symmetries \rf{16082015-man-47}-\rf{16082015-man-49}, we see the appearance of other spin operators $M^{\eta\rho}$, $M^{\rho a}$, $M^{ab}$. Our observation is as follows. If to the set of spin operators $M^{\eta a}$, $M^{\eta \eta}$, which enter the BRST operator, and the set of spin operators  $M^{\eta\rho}$, $M^{\rho a}$, $M^{ab}$, which enter the generators of the relativistic symmetries, we add the spin operator $M^{\rho\rho}$, defined by the relation $M^{\rho\rho} = \rho\rhob$, then we obtain the $osp(d-1,1|2)$  superalgebra. In other words, we deal with the following spin operators
\beq
\underbrace{ M^{ab}}_{so(d-1,1)},\, \underbrace{M^{\eta\rho}, M^{\eta\eta}, M^{\rho\rho}}_{sp(2)}\,\,,\, \underbrace{M^{\eta a}, M^{\rho a}}_{\hbox{ coset }}\,,
\eeq
which satisfy the (anti)commutation relations of the $osp(d-1,1|2)$ superalgebra,

\beq
\label{17082015-man-02} && [M^{ab},M^{ce}] = \eta^{bc} M^{ae} + 3 \hbox{ terms},
\\[5pt]
&& [M^{\eta\rho},M^{\eta \eta}] = 2 M^{\eta\eta}\,,
\nonumber\\
&& [M^{\eta\rho},M^{\rho \rho}] = - 2 M^{\rho\rho}\,,
\nonumber\\
\label{17082015-man-03} && [M^{\eta\eta},M^{\rho\rho}] = M^{\eta\rho}\,,
\\[5pt]
&& [M^{\eta\rho},M^{\eta a}] = M^{\eta a}\,,
\nonumber\\
&& [M^{\eta\rho},M^{\rho a}] = - M^{\rho a}\,,
\nonumber\\
&& [M^{\eta\eta},M^{\rho a}] = M^{\eta a}\,,
\nonumber\\
\label{17082015-man-04} && [M^{\rho\rho},M^{\eta a}] = M^{\rho a}\,,
\\[5pt]
&& [M^{\eta a},M^{bc}] = \eta^{ab} M^{\eta c} - \eta^{ac} M^{\eta b} \,,
\nonumber\\
\label{17082015-man-05} && [M^{\rho a},M^{bc}] = \eta^{ab} M^{\rho c} - \eta^{ac} M^{\rho b}\,,
\\[5pt]
&& \{M^{\rho a}, M^{\eta b}\} = \eta^{ab} M^{\eta\rho}  + M^{ab}\,,
\nonumber\\
&& \{ M^{\eta a},M^{\eta b}\} = - 2\eta^{ab} M^{\eta\eta}\,,
\nonumber\\
\label{17082015-man-06} && \{ M^{\rho a},M^{\rho b}\} =  2\eta^{ab} M^{\rho\rho}\,.
\eeq
For each  generator $X$ of the $osp(d-1,1|2)$ superalgebra, we introduce the notion of $q$-charge by using the relation $[M^{\eta\rho},X]  = q X$. Then using (anti)commutation relations \rf{17082015-man-02}-\rf{17082015-man-06}, we recall that the $osp(d-1,1|2)$  superalgebra admits the Kantor decomposition
\beq
\underbrace{M^{\rho\rho}}_{ q=-2 }\,, \  \underbrace{ M^{\rho a} }_{ q=-1 }\, \ \underbrace{M^{\eta\rho},  M^{ab}}_{ q=0 }\,, \ \underbrace{ M^{\eta a} }_{ q=1}\,, \  \underbrace{ M^{\eta\eta} }_{ q=2 }
\eeq

Fields involved into the BRST-BV formulation are classified according the representations
of the $osp(d-1,1|2)$  superalgebra. This fact is well known for fields in $R^{d-1,1}$ and it is still true for fields in $AdS_{d+1}$.
Also, it is well known that the BRST-BV formulation of fields in $R^{d-1,1}$ involves
the spin operators $M^{\eta a}$, $M^{\eta \eta}$, which enter the BRST operator, and the spin operators $M^{ab}$, which enter the generators of the Lorentz algebra $so(d-1,1)$. The spin operators $M^{\eta a}$, $M^{\eta \eta}$, $M^{ab}$ form sub-superalgebra of the $osp(d-1,1|2)$ superalgebra. From our study, we learn that, for the case of $AdS_{d+1}$, the BRST-BV formulation turns out to be more interesting.
Namely, in this case the BRST-BV formulation involves the spin operators $M^{\eta a}$, $M^{\eta \eta}$, which enter the BRST operator, and the spin operators $M^{\eta\rho}$, $M^{\rho a}$, $M^{ab}$, which enter the generators of the relativistic symmetries algebra $so(d,2)$. The set of the spin operators $M^{\eta a}$, $M^{\eta \eta}$ $M^{\eta\rho}$, $M^{\rho a}$, $M^{ab}$ do not form superalgebra. Note however, if to the just mentioned set of the spin operators we add the spin operator $M^{\rho\rho}=\rho\rhob$, which appears in anticommutator \rf{17082015-man-06}, then we get the $osp(d-1,1|2)$ superalgebra. Note also that, in the generators of the $so(d,2)$ algebra, the spin operator $M^{\rho a}$  appears together with the Grassmann coordinate $\theta$ (see \rf{16082015-man-49}). Therefore, anticommutator
\rf{17082015-man-06} does not contribute to the commutation relations of the $so(d,2)$ algebra.

In conclusion, we note that we developed the BRST-BV formulation of totally symmetric arbitrary spin massless fields in AdS space. Our approach might have interesting applications in a study of various problems of higher-spin fields in AdS space. Namely, we note the problem of the BRST-BV formulation of mixed-symmetry fields \cite{Alkalaev:2009vm}. Other interesting problem is the BRST-BV formulation of interacting higher-spin fields \cite{Boulanger:2004rx,Metsaev:2012uy,Henneaux:2013gba}. Application of  the BRST-BV approach to conformal fields in AdS space \cite{Metsaev:2014iwa} also seems to be of some interest.

\bigskip
{\bf Acknowledgments}. This work was supported by the RFBR Grant No.14-02-01171.

\setcounter{section}{0}\setcounter{subsection}{0}
\appendix{ \large Notation and conventions  }

Vector indices of the $so(d-1,1)$ algebra take values $a,b,c,e=0,1,\ldots ,d-1$. To simplify our expressions, we drop the flat metric $\eta^{ab}=(-,+,\ldots,+)$  in the scalar product :
$X^aY^a \equiv \eta_{ab}X^a Y^b$.

We use the Grassmann coordinate $\theta$, $\theta^2=0$. The left derivative for the $\theta$ is denoted as $\partial_\theta = \partial/\partial \theta$, while the integral over $\theta$ is normalized as $\int d\theta \theta =1$.
Creation operators $\alpha^a$, $\alpha^z$, $\zeta$, $\eta$, $\rho$ and the corresponding annihilation  operators $\alphab^a$, $\alphab^z$, $\zetab$, $\rhob$, $\etab$ are referred to as oscillators.
(Anti)commutation relations, the vacuum $|0\rangle $ and the hermitian conjugation rules are defined by the relations
\beq
&& [\alphab^a,\alpha^b] = \eta^{ab},  \ \quad \ [\alphab^z,\alpha^z]=1,  \ \quad \ [\zetab,\zeta]=1, \ \quad \ \{\rhob,\eta\}=1\,, \ \quad \  \{\etab,\rho\} =1\,,\qquad
\\
&& \alphab^a |0\rangle = 0\,, \hspace{1.3cm} \alphab^z |0\rangle = 0\,,  \hspace{0.9cm} \zetab |0\rangle
 = 0\,,   \hspace{1cm} \etab |0\rangle = 0\,, \hspace{1cm} \rhob |0\rangle = 0\,,
\\
&& \alpha^{a \dagger} = \alphab^a\,, \hspace{1.4cm} \alpha^{z\dagger} = \alphab^z\,, \hspace{1.1cm} \zeta^\dagger = \zetab\,, \hspace{1.3cm} \eta^\dagger = \etab\,, \hspace{1,4cm} \rho^\dagger = \rhob\,.
\eeq
The oscillators $\alpha^a$, $\alphab^a$ and $\alpha^z$, $\alphab^z$, $\zeta$, $\zetab$, $\eta$, $\rhob$, $\rho$, $\etab$ transform as the respective vectors and scalars of the Lorentz algebra $so(d-1,1)$. Derivatives of the coordinates $x^a$, $z$ are denoted by $\partial^a \equiv \eta^{ab}\partial/\partial x^b$, $\partial_z \equiv \partial/\partial z$  respectively.
We use the following notation for the products of the oscillators and the derivatives
\beq
\label{22082015-man-01} && \hspace{-1.5cm} \Box \equiv \partial^a \partial^a \, \hspace{1.6cm} \alpha^2 \equiv \alpha^a \alpha^a\,, \hspace{1.4cm} \bar\alpha^2 \equiv \bar\alpha^a \bar\alpha^a\,,
\\
\label{22082015-man-02} && \hspace{-1.5cm} N_\alpha \equiv \alpha^a \alphab^a  \,, \qquad \ \ N_z\equiv \alpha^z \alphab^z  \,, \qquad \ \ \ \  N_\zeta \equiv \zeta \zetab \,, \qquad  N_\eta \equiv \eta \rhob\,, \qquad  N_\rho \equiv \rho \etab\,.\qquad
\eeq

Hermitian conjugation rules for the coordinates and the derivatives are defined as $(x^a,z,\theta)^\dagger = (x^a,z,\theta)$, $(\partial^a, \partial_z, \partial_\theta)^\dagger  = (-\partial^a, -\partial_z,\partial_\theta)$.
Hermitian conjugation rule for the product of two operators $A$, $B$ having arbitrary ghost numbers is defined as $(AB)^\dagger = B^\dagger A^\dagger$.

The ghost numbers of $\alpha^a$, $\alpha^z$, $\zeta$ are equal to zero, the ghost numbers
of $\theta$, $\eta$, $\rhob$ are equal to 1, and the ghost numbers of $\partial_\theta$, $\etab$, $\rho$ are equal to -1.
The ghost numbers of ket-vectors in \rf{15082015-man-02},\rf{15082015-man-03} are defined as eigenvalues of the external Faddeev-Popov operator denoted by $\Nbf_\FPsm^\ext$. The eigenvalues of  $\Nbf_\FPsm^\ext$ are found from the relation $(N_\FPsm^\intrm + \Nbf_\FPsm^\ext)\Phik=0$, where the realization of the internal Faddeev-Popov ghost operator $N_\FPsm^\intrm$ on space of ket-vector $\Phik$ \rf{15082015-man-01} is given by the following relation $N_\FPsm^\intrm = \theta\partial_\theta + N_\eta - N_\rho$. Using this rule, we find the ghost numbers of ket-vectors appearing in \rf{15082015-man-02}, \rf{15082015-man-03},
\beq
\label{20082015-man-01} && \gh(|\phi_\Ism\rangle)=0\,, \hspace{1cm} \gh(|c\rangle)=1\,, \hspace{1cm} \gh(|\cb\rangle)=-1\,, \hspace{1cm} \gh(|\phi_\IIsm\rangle)=0\,,
\\
\label{20082015-man-02} && \gh(|\phi_{\Ism *}\rangle)=-1\,, \hspace{0.5cm} \gh(|c_*\rangle)=-2\,, \hspace{0.5cm} \gh(|\cb_*\rangle)=0\,,  \hspace{1.2cm} \gh(|\phi_{\IIsm *}\rangle) = -1\,.\qquad
\eeq
As the ghost numbers of the oscillators $\alpha^a$, $\alpha^z$, $\zeta$ and the vacuum $|0\rangle$ are equal to zero, the ghost numbers of the ket-vectors \rf{15082015-man-02}, \rf{15082015-man-03} coincide with ghost numbers of tensor fields which are obtained by expanding  the ket-vectors \rf{15082015-man-02}, \rf{15082015-man-03} into the oscillators $\alpha^a$, $\alpha^z$, $\zeta$. We note also, that ghost numbers of the gauge transformation parameters are found from the following relation $(N_\FPsm^\intrm + \Nbf_\FPsm^\ext+1)\Xik=0$.

Bra-vectors $\Phibr$, $\langle\phi_{\Ism,\IIsm}|$ and ket-vectors $\Phik$, $|\phi_{\Ism,\IIsm}\rangle$ are related by the following hermitian conjugation rules: $\Phibr= (\Phik)^\dagger$, $\langle\phi_{\Ism,\IIsm}| = (|\phi_{\Ism,\IIsm}\rangle)^\dagger$. Bra-vectors and ket-vectors of the Faddeev-Popov fields are related as $\cbr = (\ck)^\dagger$, $\cbbr = - (\cbk)^\dagger$, while, for the Faddeev-Popov ghost tensor fields, we use the following hermitian conjugation rules: $c^{a_1\ldots a_{s'}\dagger} = c^{a_1\ldots a_{s'}}$, $\cb^{a_1\ldots a_{s'}\dagger} = - \cb^{a_1\ldots a_{s'}}$ .

\end{document}